\documentclass[fleqn,a4paper]{article}
\usepackage{a4wide}
\usepackage{amsmath}
\usepackage{amssymb}
\usepackage{comment}
\usepackage{hyphenat}
\usepackage{multicol}
\usepackage{times}
\usepackage{bm}
\usepackage{dsfont}

\usepackage{natbib}
\usepackage{mathtools}
\usepackage{breakurl}
\usepackage[breaklinks]{hyperref}
\usepackage{dsfont}
\usepackage[caption = false]{subfig}
\usepackage{floatrow}
\newfloatcommand{capbtabbox}{table}[][\FBwidth]
\usepackage{tabularx,booktabs}
\newcolumntype{Y}{>{\centering\arraybackslash}X}

\newtheorem{theorem}{Theorem}[section]

\newtheorem{remark}[theorem]{Remark}

\newcommand{\qed}{\hfill$\Box$}

\title{Bayesian Analysis of Privacy Attacks on GPS Trajectories}
\author{Sirio Legramanti\footnote{Department of Decision Sciences, Bocconi University, 20136 Milan, Italy, \texttt{sirio.legramanti@phd.unibocconi.it}}}

\begin{document}
\maketitle

\begin{abstract}
	The success of applications for sharing GPS trajectories raises serious privacy concerns, in particular about users' home addresses.
	In this paper we show that a Bayesian approach is natural and effective for a rigorous analysis of home-identification attacks and their countermeasures, in terms of privacy. 
	We focus on a family of countermeasures named ``privacy-region strategies'', consisting in publishing each trajectory from the first exit to the last entrance from/into a privacy region.
	Their performance is studied through simulations on Brownian motions.\\
	
\noindent	\textit{Some key words:} 	
	Harmonic measure;
	Home-identification;
	Privacy protection
\end{abstract}

\section{Introduction}
\label{sec_intro}

Nowadays GPS trajectories are not only easily recorded but also massively shared, for example via sport social networks.
This raises serious privacy concerns, because from complete GPS trajectories it is possible to infer sensitive information about users, such as their home addresses \citep{liao2006location,hoh2006enhancing}.
The problem can be tackled from two sides: on one hand, users should carefully select who can access their trajectories; on the other, the amount of sensitive information contained in GPS trajectories should be limited.
Since users' privacy awareness has been shown to be unreliable \citep{krumm2009survey}, the second approach is necessary, and can be implemented via algorithms taking GPS trajectories as input and giving as output modified trajectories from which it is harder to infer sensitive information.
We call these algorithms \textit{obfuscation strategies}, but they are also known as \textit{location-privacy
	protection mechanisms} \citep{shokri2011quantifying} or \textit{trajectory privacy preservation mechanisms} \citep{singh2018choosing}. 
Obfuscation strategies range from publishing nothing to publishing original trajectories as they are. 
Both extremes are undesirable in opposite ways that highlight the tradeoff between privacy and \textit{utility}, which is the residual value of the obfuscated data for the considered application \citep{singh2018choosing}. Publishing nothing guarantees perfect privacy but null utility, while full disclosure provides maximum utility but usually insufficient privacy.

In the present work, we focus on obfuscation strategies against \textit{home-identification attacks}, in which an adversary tries to localize the house of a user exploiting his GPS trajectories.
\citet{krumm2007inference} and \citet{hoh2006enhancing} deal with the same type of attack but rely on heuristics giving as output, respectively, a single address and a list of addresses with no ordering in probability, thus failing to quantify the adversary's uncertainty, which instead we regard as a key part of a privacy measure.
To overcome this issue, we propose a Bayesian framework for assessing the efficacy of privacy attacks and their countermeasures. 
Even if implemented for home-identification attacks on GPS trajectories, our framework can be used for other types of data and privacy attacks.
A Bayesian approach is not completely new in privacy literature, but still %quite rare and used
mostly used to formulate privacy definitions \citep{realisticadversaries,bassily2013coupled,semantics}.
A partial exception can be found in  
\citet{shokri2011quantifying}, who however
employ posterior bias as a privacy measure, completely disregarding the uncertainty quantification that naturally comes with a Bayesian approach.

\section{Framework for Home-Identification Attacks}
\label{sec_framework}

We model home-identification attacks on GPS trajectories as Bayesian inference problems in which the parameter of interest is the user's house location $ {\theta \in \Theta \subseteq \mathds{R}^2} $ and data consist of $ n $ published GPS trajectories $ y^{(i)} = \{y_t^{(i)}\}_{t \in [0,\widetilde{T}^{(i)}]} $ $ (i=1,\ldots,n) $, possibly altered before publication.
We assume that the adversary has no access to the corresponding original trajectories $ x^{(i)} = \{x_t^{(i)}\}_{t \in [0,T^{(i)}]} \ (i=1,\ldots,n)$, but knows the adopted obfuscation strategy and has some background knowledge about the user's house location $ \theta $, modeled by a prior distribution $ \pi $.
Under obfuscation strategy $ s $ and model $ m $ for the original trajectories, the published trajectories are then described by the hierarchical model
\begin{equation*}
\label{eq_strategy}
(y^{(i)} \mid x^{(i)}, \theta) \sim p_s, \qquad
(x^{(i)} \mid \theta) \sim p_m, \qquad 
\theta \sim \pi \qquad 
(i=1,\ldots,n).
\end{equation*}
We model both original and published GPS trajectories with continuous time stochastic processes even if GPS sensors actually record the position at discrete times, because the sampling frequency is usually high (a typical value is 1 Hz).

\subsection{Privacy and Utility Measures}
\label{subsec_measures}

Since we model privacy attacks as point estimation problems, it is natural to measure privacy through the quality of the adversary's estimate: the better the estimate, the poorer the privacy.
This principle holds for any privacy attack, while the type of sensitive attribute determines the loss function used to evaluate the estimate.
Since in our case the sensitive attribute is the user's house location $ \theta \in \Theta \subseteq \mathds{R}^2 $, we adopt a quadratic loss function. We then employ the posterior Mean Square Error (MSE) as a measure of the quality of the adversary's estimate, and hence as a measure of privacy.
Through its well-known bias-variance decomposition, MSE incorporates both \textit{correctness} and \textit{uncertainty}, which are discussed as privacy measures in \citet{shokri2011quantifying}.

While the privacy measure depends on the sensitive attribute, the utility measure depends on the considered application.
In the present work we focus on sharing GPS trajectories of fitness activities via sport social networks. 
In this context, fake times or locations are not allowed, since they would alter rankings. 
The only admitted form of perturbation is cutting part of the original trajectories.
In particular, users may accept cuts to the initial and final part of their trajectories, which usually consist of how they get out of their neighborhood and back in.
However, we assume that users want these cuts to be as small as possible.
Based on these assumptions, the utility of a published trajectory $ y = \{ y_t \}_{t \in [0,\widetilde{T}]}$ with respect to the corresponding original trajectory $ x = \{ x_t \}_{t \in [0,T]}$ can be defined as a monotone-decreasing function $ \phi $ of the following \textit{squared perturbation} (SP):
\begin{eqnarray}\label{eq_sp}
SP(y,x) = \left\{ 
\begin{array}{ll}
\| y_0 - x_0 \|^2 + \| y_{\widetilde{T}} - x_{T} \|^2 & \quad \text{if \ $ y = x_{\big|[t_1,t_2]} $, with $ 0 \leq t_1 \leq t_2 \leq T $}\\
\infty & \quad \text{otherwise}\\
\end{array} \right. 
\label{eq_discr_2}
\end{eqnarray}
where $x_{\big|[t_1,t_2]} $ denotes the restriction of $ x $ to $ [t_1,t_2] $ and $ \| \cdot \| $ the Euclidean norm in $ \mathds{R}^2 $.
If the published trajectory is obtained by cutting the initial and/or final part of the original one, the SP is the sum of the square Euclidean distances between their starting and ending points, otherwise it is set to infinity, leading to the lowest possible utility.

The variability of original trajectories and the stochasticity of obfuscation strategies make 
the SP random. 
We then define the utility of obfuscation strategy $s$, under model $ m $ for original trajectories and prior $\pi$ on $ \theta $, as
\begin{equation} \label{eq_def_utility} 
U_{m,\pi}(s) = \int \phi(SP(y,x)) 
\ dp_s(y \mid x,\theta) \ dp_m (x \mid \theta)
\ d\pi(\theta). 
\end{equation}
In the present work, we do not need to quantify utility but just to set the same level of utility to fairly compare two strategies, hence we do not need to specify $\phi$.
A sufficient condition for two strategies to have the same utility, for any choice of $ \phi $, is to induce the same SP distribution.

\subsection{Privacy-Region Strategies}
\label{subsec_strategies}

As stated in \S\ref{subsec_measures}, the only obfuscation strategies that are admissible for fitness GPS trajectories are those cutting at most the initial and final part of original trajectories.
Since most of fitness activities either start and finish close to the user's house location $ \theta $ or do not come close to it at all, this can be achieved by a class of strategies, that we call \textit{privacy-region strategies}, 
cutting the part of each original trajectory before the first exit and after the last entrance from/into a privacy region containing $ \theta $. 
Each trajectory may be cut using a different privacy region, hence we denote with $ D_i $ the privacy region for the  $ i $-th trajectory.
A privacy-region strategy $ s $ is described by 
\begin{equation*}
(y^{(i)} \mid x^{(i)}, D_i) = x^{(i)}_{\big|[t^{(i)}_1,t^{(i)}_2 ]}, \qquad 
(D_i \mid x^{(i)},\theta) \sim p_s \qquad (i=1,\ldots,n)
\end{equation*} 
with $ t^{(i)}_1 = \inf \{t \in [0,T^{(i)}]: x_t^{(i)} \notin D_i \} $ and $ t^{(i)}_2 =
\sup \{t \in [0,T^{(i)}]: x_t^{(i)} \notin D_i \} $. If the infimum for $ t^{(i)}_1 $ is over an empty set (i.e. the original trajectory never leaves the privacy region) then no trajectory is published and the SP is set to infinity.

The distribution of each privacy region $ D_i $ 
may depend on the corresponding original trajectory, allowing to design adaptive obfuscation strategies. In this work, though, we do not implement this dependency, leaving it to future research.

We now describe two privacy-region strategies, corresponding to different distributions of the privacy regions: \textit{random-radius strategy} and \textit{two-balls strategy}.

\begin{figure} 
	\centering
	\subfloat[]{\includegraphics[width=0.37\columnwidth]{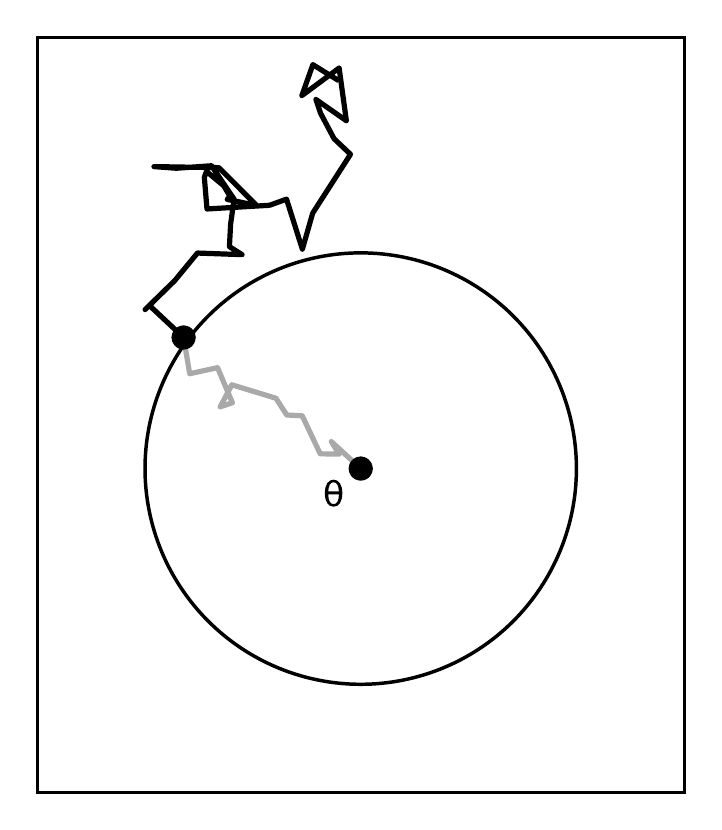}\label{subfig_random_radius}} \hspace{0.05\columnwidth}
	\subfloat[]{\includegraphics[width=0.37\columnwidth]{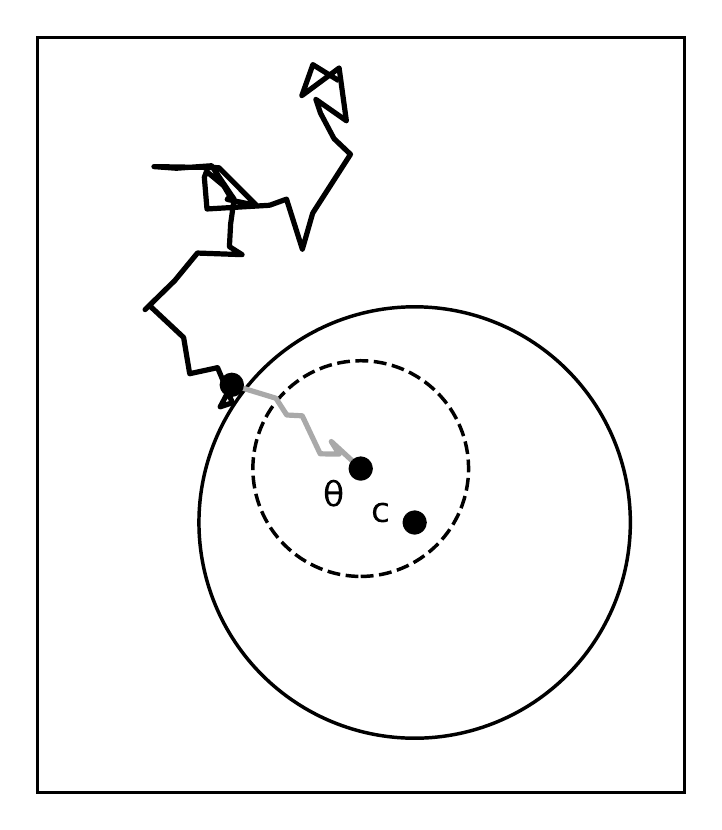}\label{subfig_two_balls}}
	\caption{
		Random-radius strategy \protect\subref{subfig_random_radius} and two-balls strategy \protect\subref{subfig_two_balls} on the same original trajectory.
		The unpublished part of the trajectory is plotted in grey, while the published part is in black. The privacy balls have solid border: for the random-radius strategy the privacy ball is centered in $ \theta $, while for the two-balls strategy it is centered in a random point $ c $ within the ball with dashed border.
	}
	\label{fig_strategies}
\end{figure}

In a random-radius strategy of parameters $ \alpha > 0 $ and $\beta > 0 $, the privacy regions are balls centered in the user's house location $ \theta $, with i.i.d. random radii $ r_i $ such that $ r_i^2 \sim Gamma(\alpha,\beta) $ (see Figure \ref{subfig_random_radius}).
\begin{remark} \label{rem_fixed}
	Other choices for the distribution of $ r_i $ are possible. The case $ r_i \sim \delta_{r^*} $, with $r^* > 0$ fixed, leads to \textit{fixed-radius strategy}.
	We will not analyze this strategy, even if used in industry, since an adversary knowing	$ r^* $ can locate $ \theta $ with no uncertainty based on just three distinct exit/entrance points from the privacy region, which in this case is almost surely unique, leveraging elementary geometry.
\end{remark}

In a two-balls strategy of parameters $ \alpha > 0 $, $ \beta > 0 $, $ r > 0$ and $ R > r $,  all the trajectories of the same user are cut employing as a privacy region the same ball $ B(c,R) $, where the radius $ R $ is fixed while the center $ c $ is randomly distributed within $ B(\theta,r) $ as follows: 
$ c = \theta + r \rho [ \cos \tau, \sin \tau]^T$, $\tau \sim U(0,2\pi)$, $\rho^2 \sim Beta(\alpha,\beta) $ (see Figure \ref{subfig_two_balls}).
This strategy generalizes the one proposed in \citep{krumm2007inference}, where $ c $ was uniformly distributed within $ B(\theta,r) $, corresponding to $ \alpha = \beta = 1 $.
The restriction $ r < R $ ensures that $ \theta $ lies within the privacy  ball.
A value of $ r $ too close to zero, though, would almost reduce the two-balls strategy to the fixed-radius strategy, that we have shown to be inefficient in Remark \ref{rem_fixed}.
Also $ \rho^2 $ concentrated near zero would produce the same effect.
On the other extreme, $ \rho^2 $ concentrated near one corresponds to $ c $ concentrated near $ \partial B(\theta,r) $. In this case, if moreover $ r $ is close to $ R $, we have that $ \theta $ will be close to the boundary of the privacy ball with high probability, producing exit points concentrated close to $ \theta $.
Hence, small values of $ r $ and distributions of $ \rho^2 $ concentrated near zero or one should be avoided.

\subsection{Privacy-Region Strategies on Brownian Motions}
\label{subsec_brownian_motion}

Our framework poses no limitation to the complexity of the model for the original trajectories, but we start with one of the simplest possible models and assume that, given the user's house location $\theta$, the original trajectories are independent Brownian motions starting at $\theta$.
Even if this is not a realistic model for human movement, it can be useful to rule out candidate obfuscation strategies. 
In fact, if a strategy is unable to hide the start of a Brownian motion, it cannot aim to hide the start of the much more structured human movement.

Thanks to the memoryless property of the Brownian motion, we can consider a simplified version of privacy-region strategies, consisting in cutting just the part of each original trajectory before the first exit from the corresponding privacy region:
\begin{equation*}
(y^{(i)} \mid x^{(i)}, D_i) =
x^{(i)}_{\big|[t^{(i)},+\infty)}, 
\qquad  (D_i \mid \theta) \sim p_s 
\qquad (i=1,\ldots,n),
\end{equation*}
with $ t^{(i)} = \inf \{t \in \mathds{R}^+: x_t^{(i)} \notin D_i \} $.
The squared perturbation \eqref{eq_sp} also simplifies to $ \| y_0 - x_0 \|^2 $.
The exit time $ t^{(i)} $ is a stopping time for $\{x_t^{(i)}\}_{t \in \mathds{R}^+}$ \citep[Remark 2.14]{morters2010brownian} and, if the privacy region $ D_i $ is almost surely a bounded domain containing $\theta$, then $ t^{(i)} $ is almost surely finite \citep[Chapter 4.3]{chung2013lectures}. Moreover,
conditional on the exit point of the original trajectory from $ D_i $, the published part of the trajectory is a Brownian motion independent of the unpublished part \citep[Theorem 2.16]{morters2010brownian}. 
It follows that the exit points of the original trajectories from the corresponding privacy regions are sufficient statistics for $\theta$. Given $ D_i $ and $ \theta $, each exit point $ z_i \ (i=1,\ldots,n) $ is distributed according to the harmonic measure of parameter $ \theta $ on the boundary of $ D_i $ \citep{kakutani1944}, a distribution that we denote with $\mathcal{H}_\theta^{D_i}$.

In conclusion, if the original trajectories are Brownian motions starting at $ \theta $, home-identification attacks against privacy-region strategies can be represented as Bayesian inference problems with the following hierarchical structure:
\begin{equation*}
(z_i \mid D_i, \theta) \sim \mathcal{H}_\theta^{D_i}, 
\qquad (D_i \mid \theta) \sim p_s, 
\qquad \theta \sim \pi 
\qquad (i=1,\ldots,n).
\end{equation*}

\section{Simulation Studies}
\label{sec_simulations}

We consider illustrative simulations on Brownian motions to understand which strategy among random-radius and two-balls guarantees higher privacy. 
The comparison is fair only if the two strategies have the same utility \citep{singh2018choosing}. 
Under our utility definition \eqref{eq_def_utility}, a sufficient condition for this is having the same SP distribution.
We relax this condition and constrain just the first two SP moments to be the same. 
Such moments are explicitly available only for the random-radius strategy, for which $ SP \sim Gamma(\alpha,\beta) $. 
Hence, we fix the two-balls parameters first, then set the random-radius ones so that the first two theoretical SP moments of the random-radius strategy match the first two sample SP moments of the two-balls strategy.

\begin{figure}
	\begin{floatrow}
		\capbtabbox{%
			\begin{tabular}{p{0.3cm}p{0.3cm}p{0.3cm}p{0.3cm}p{1.2cm}p{0.8cm}p{0.8cm}}
				\multicolumn{4}{c}{TB parameters} & & \multicolumn{2}{c}{MSE median} \\
				r & R & $ \alpha $ & $ \beta $ & mean~SP & TB & RR \\ \toprule
				1 & 3 & 4 & 4 & 8.34 & 0.25 & 0.03 \\
				1 & 4 & 4 & 4 & 15.34 & 0.48 & 0.04 \\
				2 & 5 & 4 & 4 & 23.23 & 0.69 & 0.13 \\
				1 & 5 & 4 & 2 & 24.29 & 0.69 & 0.05 \\
				1 & 5 & 4 & 4 & 24.42 & 0.78 & 0.04 \\
				1 & 5 & 2 & 4 & 24.71 & 0.77 & 0.02 \\
			\end{tabular} 
		}{%
			\caption{Settings considered for the comparison between the two-balls strategy (TB) and the random-radius strategy (RR).}%
			\label{tab_settings}
		}
		\ffigbox{%
			\includegraphics[width=\columnwidth]{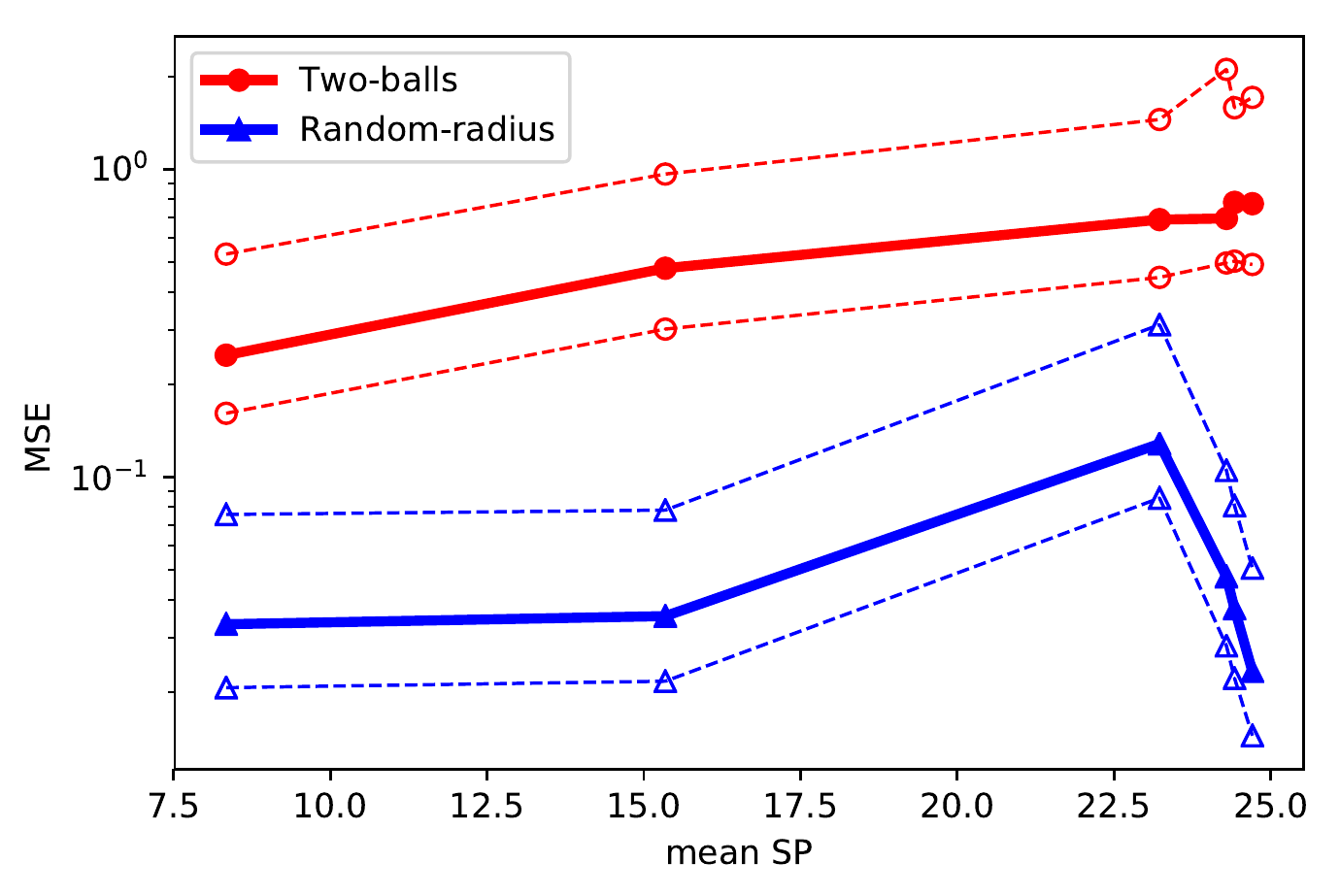} 
		}{%
			\caption{Posterior MSE given 50 trajectories for the settings in Table \ref{tab_settings}. The median is plotted solid, while 5\% and 95\% quantiles are dashed.}%
			\label{fig_scatter}
		}
	\end{floatrow}
\end{figure}

We consider different sets of two-balls parameters, listed in Table \ref{tab_settings}.
For each setting, we generate 50 Brownian motion trajectories starting at $ \theta $, which is fixed at the origin since the considered strategies are translation invariant.
We then process these trajectories through both strategies, producing 50 cut trajectories each, on which we perform
Bayesian inference for $ \theta $, under a uniform improper prior, using \textit{PyStan} \citep{pystan}.
From the posterior samples of $ \theta $ we compute a Monte Carlo estimate of the posterior MSE, plotted in Figure \ref{fig_scatter} for all settings in Table \ref{tab_settings}. 
We can observe that the MSE, which is our measure of privacy, is higher with the two-balls strategy than with the random-radius strategy, in all settings. 
Especially with the two-balls strategy, higher SP (i.e. lower utility) corresponds to higher MSE (i.e. higher privacy), in the expected privacy-utility tradeoff.

\begin{figure} 
	\centering
	\includegraphics[width=0.65\columnwidth]{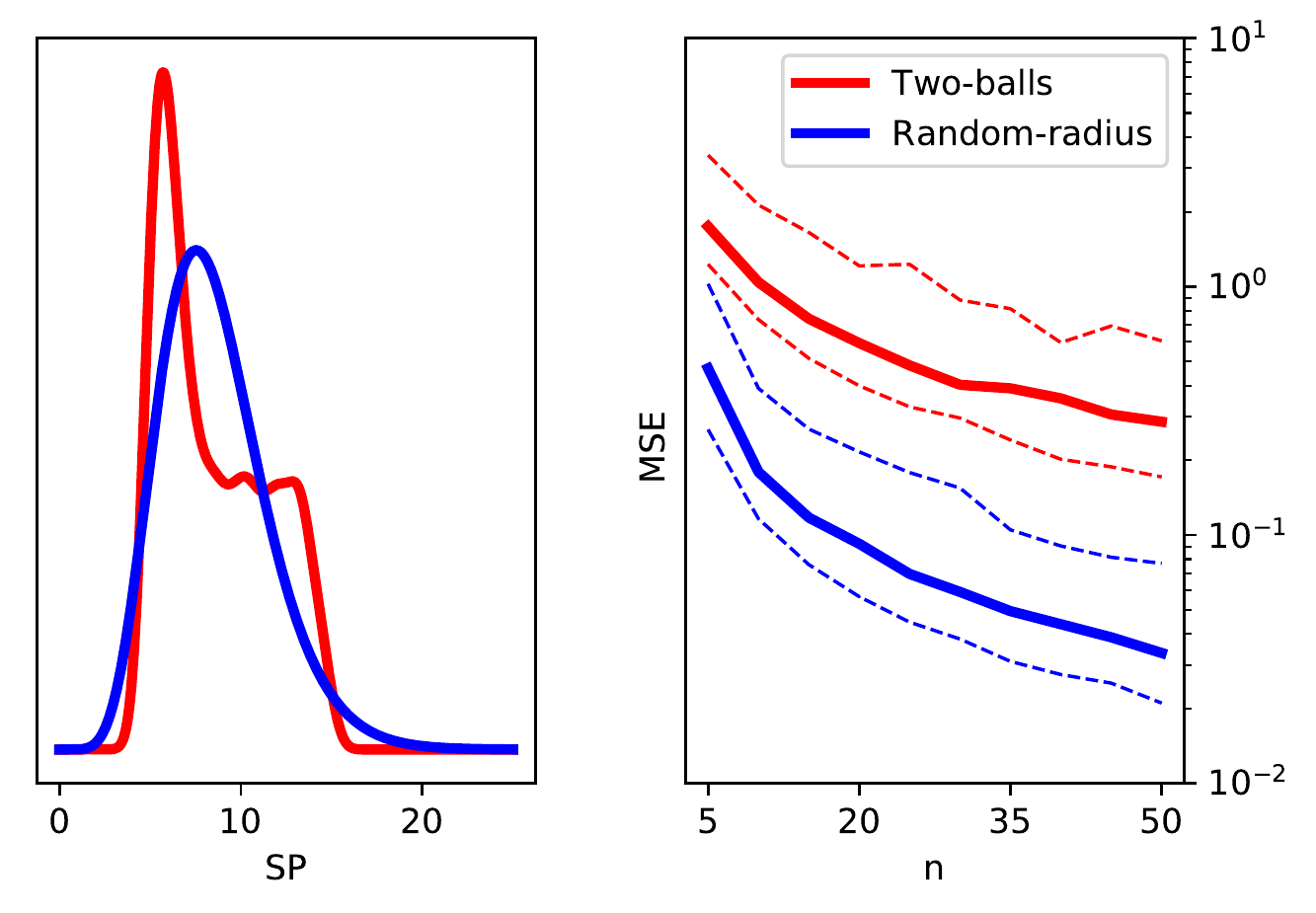}
	\caption{Two-balls strategy vs. random-radius strategy as sample size grows, for a single set of two-balls parameters ($ r=1 $, $ R=3 $, $ \alpha=\beta=4 $). On the left, the SP pdf; on the right, the posterior MSE median (solid) and its 5\% and 95\% quantiles (dashed) as functions of the sample size $ n $.}	
	\label{fig_compare}
\end{figure}

In Figure \ref{fig_compare} we plot the posterior MSE as sample size grows, for a single setting ($ r=1 $, $ R=3 $, $ \alpha=\beta=4 $).
The MSE goes to zero under both strategies, but is stably higher with the two-balls strategy than with the random-radius strategy.
We can also observe that matching the first two SP moments produces SP densities that are similar in shape as well, making the comparison particularly fair. 

Finally note that each home-identification attack on 50 trajectories took about one second on a regular laptop (Intel Core i7-3632QM CPU @ 2.20GHz x 8, 7.7 GB of RAM), with running time scaling linearly in the number of trajectories.
The modest amount of time and computational resources needed for the attack confirms that risks for users' privacy are real.

\section{Discussion}
\label{sec_discussion}

Further analysis of privacy-region strategies may involve the development of more realistic models for GPS trajectories, which can still be plugged in our framework. This may lead to the loss of some of the properties exploited here, as the sufficiency of exit points or the availability of their likelihood. Anyway, as long as a generative model and some quasi-sufficient statistics are available, 
Bayesian inference can still be carried out
through Approximate Bayesian Computation \citep{beaumont2002approximate}. 
Exit points, which are sufficient statistics in the Brownian motion case, are the first candidates as quasi-sufficient statistics under more general models.

Other possible extensions of the present work 
are represented by different applications, obfuscation strategies or even data-types. In fact, the utility measure is the only application-specific element in our framework, which in all the other aspects is completely general.

\section*{Acknowledgments}
I would like to thank Giacomo Aletti, Daniele Durante, Nelly Litvak and Lucia Paci for their helpful comments on a first version of this work.

\bibliographystyle{apalike}
\bibliography{privacy_SIS}

\begin{thebibliography}{}

\bibitem[Bassily et~al., 2013]{bassily2013coupled}
Bassily, R., Groce, A., Katz, J., and Smith, A. (2013).
\newblock Coupled-worlds privacy: Exploiting adversarial uncertainty in
  statistical data privacy.
\newblock In {\em 2013 IEEE 54th Annual Symp. on Found. of Computer Science},
  pages 439--448.

\bibitem[Beaumont et~al., 2002]{beaumont2002approximate}
Beaumont, M.~A., Zhang, W., and Balding, D.~J. (2002).
\newblock Approximate {Bayesian} computation in population genetics.
\newblock {\em Genetics}, 162(4):2025--2035.

\bibitem[Chung, 2013]{chung2013lectures}
Chung, K.~L. (2013).
\newblock {\em Lectures from {Markov} processes to {Brownian} motion}, volume
  249.
\newblock Springer.

\bibitem[Hoh et~al., 2006]{hoh2006enhancing}
Hoh, B., Gruteser, M., Xiong, H., and Alrabady, A. (2006).
\newblock Enhancing security and privacy in traffic-monitoring systems.
\newblock {\em Pervasive Computing}, 5(4):38--46.

\bibitem[Kakutani, 1944]{kakutani1944}
Kakutani, S. (1944).
\newblock Two-dimensional {Brownian} motion and harmonic functions.
\newblock {\em Proc. of the Imp. Acad.}, 20(10):706--714.

\bibitem[Kasiviswanathan and Smith, 2014]{semantics}
Kasiviswanathan, S.~P. and Smith, A. (2014).
\newblock On the 'semantics' of differential privacy: A {Bayesian} formulation.
\newblock {\em J. of Privacy and Confidentiality}, 6(1):1--16.

\bibitem[Krumm, 2007]{krumm2007inference}
Krumm, J. (2007).
\newblock Inference attacks on location tracks.
\newblock In {\em Int. Conf. on Pervasive Computing}, pages 127--143. Springer.

\bibitem[Krumm, 2009]{krumm2009survey}
Krumm, J. (2009).
\newblock A survey of computational location privacy.
\newblock {\em Personal and Ubiquitous Computing}, 13(6):391--399.

\bibitem[Liao et~al., 2006]{liao2006location}
Liao, L., Fox, D., and Kautz, H. (2006).
\newblock Location-based activity recognition.
\newblock In {\em Adv. in Neural Information Processing Systems}, pages
  787--794.

\bibitem[Machanavajjhala et~al., 2009]{realisticadversaries}
Machanavajjhala, A., Gehrke, J., and G{\"o}tz, M. (2009).
\newblock Data publishing against realistic adversaries.
\newblock {\em Proc. of the VLDB Endowment}, 2(1):790--801.

\bibitem[M{\"o}rters and Peres, 2010]{morters2010brownian}
M{\"o}rters, P. and Peres, Y. (2010).
\newblock {\em Brownian motion}.
\newblock Cambridge Univ. Press.

\bibitem[Shokri et~al., 2011]{shokri2011quantifying}
Shokri, R., Theodorakopoulos, G., Le~Boudec, J.-Y., and Hubaux, J.-P. (2011).
\newblock Quantifying location privacy.
\newblock In {\em 2011 {IEEE} Symp. on Security and Privacy}, pages 247--262.

\bibitem[Singh et~al., 2018]{singh2018choosing}
Singh, R., Theodorakopoulos, G., Marina, M.~K., and Arapinis, M. (2018).
\newblock On choosing between privacy preservation mechanisms for mobile
  trajectory data sharing.
\newblock In {\em 2018 IEEE Conf. on Commun. and Network Security}, pages 1--9.

\bibitem[{Stan Development Team}, 2017]{pystan}
{Stan Development Team} (2017).
\newblock Pystan: the {Python} interface to {Stan}, {Version} 2.16.0.0.
\newblock \url{http://mc-stan.org}.
\newblock Online: accessed March 12, 2019.

\end{thebibliography}

\end{document}